\documentclass[letterpaper,twocolumn,10pt]{article}
\usepackage{booktabs}
\usepackage{hyperref}
\usepackage{usenix,epsfig,endnotes}
\usepackage[utf8]{inputenc}
\usepackage[T1]{fontenc}
\usepackage{lineno}
\usepackage{bbding}
\usepackage{array}
\usepackage{graphicx}
\usepackage{epstopdf}
\usepackage{multicol}
\usepackage{multirow}
\usepackage{float}
\usepackage{subfig}

\begin{document}

\date{}

\title{\Large \bf Real-CATS: A Practical Training Ground for Emerging Research \\on Cryptocurrency Cybercrime Detection}

\author{
{\rm Jiadong Shi}\\
Southeast University
\and
{\rm Chunyu Duan}\\
Southeast University
\and
{\rm Hao Lei}\\
Southeast University
\and
{\rm Liangmin Wang*}\\
liangmin@seu.edu.cn\\
Southeast University
}

\maketitle

\thispagestyle{empty}

\subsection*{Abstract}
   Cybercriminals pose a significant threat to blockchain trading security, causing \$40.9 billion in losses in 2024. However, the lack of an effective real-world address dataset hinders the advancement of cybercrime detection research. The anti-cybercrime efforts of researchers from broader fields, such as statistics and artificial intelligence, are blocked by data scarcity. In this paper, we present Real-CATS, a Real-world dataset of Cryptocurrency Addresses with Transaction profileS, serving as a practical training ground for developing and assessing detection methods. Real-CATS comprises 103,203 criminal addresses from real-world reports and 106,196 benign addresses from exchange customers. It satifies the C3R characteristics (Comprehensiveness, Classifiability, Customizability, and Real-world Transferability), which are fundemental for practical detection of cryptocurrency cybercrime. The dataset provides three main functions: 1) effective evaluation of detection methods, 2) support for feature extensions, and 3) a new evaluation scenario for real-world deployment. Real-CATS also offers opportunities to expand cybercrime measurement studies. It is particularly beneficial for researchers without cryptocurrency-related knowledge to engage in this emerging research field. We hope that studies on cryptocurrency cybercrime detection will be promoted by an increasing number of cross-disciplinary researchers drawn to this versatile data platform. All datasets are available at \url{https://github.com/sjdseu/Real-CATS}.

\section{Introduction}

Since its inception in 2008~\cite{nakamoto2008bitcoin}, Bitcoin has become increasingly prevalent in the financial and commercial fields. The Bitcoin system provides pseudo-anonymity for legitimate users to protect transaction privacy, which has also attracted cybercriminals to abuse cryptocurrencies as a method of payment, such as ransomware~\cite{CONTI2018162,Huang2018SP,PC2019ransom}, sextortion~\cite{oggier2020ego}, human trafficking~\cite{Portnoff2017human} and blackmailing~\cite{Paquet-Clouston2019AFT, Bartoletti2021CryptocurrencySA}. With the rising popularity and appreciation of crypto assets, cybercriminals are attracted to directly steal cryptocurrencies~\cite{He2023Txphish,Li2023DoubleAN,Liu2024Fish}. Likewise, Ethereum is heavily abused for scams and money laundering due to its diverse ecosystem~\cite{Chen2021ijcai, Du2024ethmix,Li2023SIEGE}. Some phishing scams~\cite{Li2023DoubleAN} operate across double-chain, simultaneously stealing the two most popular cryptocurrencies: Bitcoin (BTC) and Ether (ETH). Based on the crypto crime report of Chainalysis~\cite{chainalysis}, more than \$40.9 billion flowed to illicit addresses in 2024, reflecting the harms that cybercrime poses to cryptocurrency finance. 

Blockchain-based cryptocurrency uses a string of characters (i.e. address) as the wallet. Cybercriminals can create multiple temporary addresses (i.e., criminal addresses) to collect illicit revenues and to launder money. By analyzing these criminal addresses, researchers can gain insights into the behavioral patterns of cybercrime activities and, consequently, help developers enhance the trading security of the blockchain ecosystem~\cite{CONTI2018162,Huang2018SP,Paquet-Clouston2019AFT,Li2023DoubleAN,He2023Txphish,Liu2024Fish,lee2019cybercriminal,Portnoff2017human,Christin2013Silk,Ron2014Dread,dosDarknetmarket2024,oggier2020ego,Bartoletti2021CryptocurrencySA,Chen2021ijcai,Du2024ethmix,Li2023SIEGE,PC2019ransom}.

In recent years, researchers have resorted to machine learning methods to detect cryptocurrency cybercrime on Bitcoin and Ethereum by analyzing transaction behavior. These efforts can be roughly categorized into two groups: feature engineering-based methods~\cite{Bartoletti2018ponziDetect,Alarab2020comparativeML} and node embedding-based methods~\cite{Oliveira2021GuiltyWalker,Wu2022Trans2vec,trans2vecVar1,trans2vecVar2,Alarab2023GCNBit,Li2022TTAGN,Li2023SIEGE}. Feature engineering-based methods typically extract features such as the transaction number and lifespan of an address, and select a proper traditional machine learning model to identify the criminal addresses. For instance, Bartoletti et al.~\cite{Bartoletti2018ponziDetect} extract 11 features including lifetime, activity days, and Gini coefficient, utilizing three machine learning algorithms to detect Bitcoin addresses related to Ponzi schemes. Node embedding-based methods use random walk or graph neural networks on the transaction graph to automatically generate the embedding representation of an address, and subsequently, a classifier network or traditional machine learning model is trained to predict the label of target addresses. For example, Trans2vec~\cite{Wu2022Trans2vec} designs biased random walk strategies based on transaction amount and timestamp and performs random walks to obtain the node embeddings. Then, a one-class SVM is applied to detect criminal addresses.

Although these methods demonstrate strong detection capabilities, the lack of an effective real-world dataset leads to evaluation results that are still incomparable or unreliable. It also prevents researchers without cryptocurrency-related knowledge, such as statistics and artificial intelligence, from contributing to this work. Specifically, research conducted on Ethereum typically utilizes privately collected data to evaluate their algorithms~\cite{Liu2024Fish,Li2022TTAGN,Li2023SIEGE}. Although these studies specify the sources of data collection, they do not provide specific criminal and benign addresses, making it difficult to compare experimental results with those of other studies. Even when some studies disclose the criminal addresses used~\cite{Chen2021ijcai}, the evaluation results remain incomparable, as the selection of benign addresses, which remain undisclosed, also significantly influences the results. Moreover, the scarcity of labeled addresses~\cite{Chen2021ijcai,Liu2024Fish,Li2022TTAGN,Li2023SIEGE,Wu2022Trans2vec,Yang2024DynEth} may impact their practicality, as we found that more than 3,000 criminal addresses are active on Ethereum on the same day. Additionally, some studies share transaction graph data~\cite{Wu2022Trans2vec}, but the information provided is only a small subset of the original transaction data, limiting the potential for customization and innovation. Likewise, some research on Bitcoin utilizes publicly available datasets with practicality limitations that affect model practicality, rendering evaluation results unreliable. For example, the labeling process of the widely used Elliptic~\cite{weber2019elliptic} dataset is informed by a heuristic-based reasoning process, which is too pure to cover real-world behavior of criminal addresses, and may harm its transferability to the real world.

To tackle these problems, we presents Real-CATS, a Real-world dataset of Cryptocurrency Addresses with Transaction profileS, to provide a practical training ground for developing and assessing detection methods. Generally, Real-CATS comprises include 90,612 and 12,591 criminal addresses on Bitcoin and  Ethereum, derived from real abuse reports submitted by victims. For benign addresses, a set of 90,176 Bitcoin addresses and 16,020 Ethereum addresses is collected from the output of the exchange hot wallet. Each address has a profile including 32 and 52 features to describe Bitcoin and Ethereum addresses, respectively. Ethereum address profiles further comprise interaction behaviors. Moreover, we collect a supplementary testing dataset, Sup-CATS, to test the comprehensiveness of our Real-CATS, including 3,147 criminal and 11,058 benign addresses on Ethereum, which simulate a real-world deployment scenario. Real-CATS dataset satisfies C3R characteristics as follows:

\begin{itemize}
    \item \textbf{Comprehensiveness}. This dataset covers more possible distributions of both criminal and benign behaviors to ensure the model's generalization ability. As far as we know, this is the largest labeled dataset for cryptocurrency cybercrime detection research. 
    
    \item \textbf{Classifiability}. This dataset contains a rich diversity of informative features that can reflect the differences between criminal and benign addresses. Moreover, compared to existing datasets, patterns in Real-CATS do not exhibit overly apparent distinguishability.

    \item \textbf{Customizability}.This dataset provides both statistical transaction profiles and detailed transaction records, allowing for customization. The transaction profile reflect the transaction behavior of an address, helping users intuitively understand the characteristics of an address. However, transaction records are more crucial as they offer researchers the flexibility to customize the data and foster the discovery of new features and innovative methods.
    
    \item \textbf{Real-world Transferability}. This dataset encompasses real-world noise, and provides a testing dataset to simulate real-world scenario, ensuring that conclusions drawn and models trained on this dataset can be applied in the real world. 
\end{itemize}

With these characteristics, Real-CATS offers a versatile platform for cryptocurrency cybercrime detection research. It also supports the feature customization, such as extracting statistical features or constructing attributed graphs. Additionally, it provides a testing dataset to assess the model's real-world transferability, thereby broadening the evaluation scope of existing research. Furthermore, this dataset can serve as seed addresses to expand other cybercrime measurement studies~\cite{Gomez2022watch,Wu2024assetflow,Gomez2023Estimate}. Overall, Real-CATS can facilitate the involvement of researchers from diverse fields in cybercrime detection, without the need for in-depth knowledge of cryptocurrency and blockchain.

To sum up, the contributions of this paper can be summarized as follows:

\begin{itemize}
    \item We propose the Real-CATS dataset, the first practical training ground for emerging research on cybercrime detection. This dataset offers an opportunity for cross-discipline researchers, such as statistics and artificial intelligence, to contribute to cryptocurrency cybercrime detection.
    
    \item This dataset satisfies the C3R characteristics (Comprehensiveness, Classifiability, Customizability, and Real-world Transferability). These characteristics are fundamental for practical detection of cryptocurrency cybercrime, facilitating the proposal of new features or algorithms by researchers.
    
    \item Real-CATS offers a versatile platform to support diverse tests for anti-cybercrime research. Its functions include effective evaluation and comparison of cybercrime detection methods, feature customization, assessment of the model's real-world transferability, and serving as seed addresses to expand other cybercrime measurement studies.

    \item The Real-CATS dataset comprises 103,203 criminal addresses and 106,196 benign addresses on Bitcoin and Ethereum, with statistical transaction profiles and detailed transaction records. And we collect a supplementary dataset Sup-CATS, including 3,147 criminal and 11,058 benign addresses from Ethereum. The dataset provides robust support for cybercrime detection research and contribute to efforts in combating cryptocurrency cybercrime.
\end{itemize}

\section{Background}
In this section, we introduce the transaction-related concepts of Bitcoin and Ethereum. We will also briefly review the existing cybercrime detection methods on both blockchains.

\subsection{Bitcoin}
Bitcoin is a transaction-based ledger where each block in the blockchain contains transaction data such as input, output, amount of Bitcoin (BTC), and transaction fee. The system relies on Unspent Transaction Outputs (UTXOs), which are used as inputs for new transactions. We detail the related concepts as follows:

\noindent \textbf{Address}. 
An address is typically represented as a unique identifier consisting of alphanumeric characters. Each address is controlled by a public key and a private key, which are used to receive and send BTC. The cost of generating an address is virtually zero due to minimal cryptographic calculations, offline generation, and the availability of free tools. As a result, cybercriminals often take advantage of this to launder money.

\noindent \textbf{UTXO}. A UTXO (Unspent Transaction Output) represents the unused output of a previous transaction that serves as input for a new transaction. Each UTXO is linked to a public key and can only be spent using the corresponding private key, meaning it is managed by a specific address. A UTXO can only be used once and cannot be directly split.

\noindent \textbf{Transaction}. A Bitcoin transaction utilizes one or more Unspent Transaction Outputs (UTXOs) from prior transactions as inputs to generate new outputs. Each input identifies a specific UTXO through its transaction hash and output index, accompanied by a cryptographic signature to verify ownership. The transaction defines new outputs, detailing recipient addresses and the amounts to transfer. Any surplus input value beyond the required outputs is returned to the sender as a "change" output. The total input value must equal the sum of the outputs and a small transaction fee. After creation, the transaction is broadcast to the Bitcoin network, validated, and eventually included in a block on the blockchain, ensuring its transparency and irreversibility.

\subsection{Ethereum}
Ethereum is an account-based blockchain system, where each account holds a certain amount of Ether (ETH) and can perform transactions. It supports both peer-to-peer transactions and smart contracts, which allow executing transactions with predefined transaction logic. We introduce related concepts as follows:

\noindent \textbf{Account}. Ethereum accounts can be categorized into two types: Externally Owned Accounts (EOAs) and Contract Accounts (CAs). EOAs are controlled by private keys, which allow users to execute transactions. CAs are controlled by smart contracts and managed by contract codes instead of private keys. CAs can only respond to transactions and calls initiated by other accounts. Both types of accounts can hold ETH and tokens. Note that each account has a unique address string as the identifier so we also use "address" to refer to "account" in this paper.

\noindent \textbf{Transaction}. An Ethereum transaction is a message signed with the private key of an account, which can be used to transfer ETH to other accounts, invoke functions of smart contracts, or deploy smart contracts. Each transaction has only one sender and receiver and requires a fee paid in Gas, with the Gas price and limit determined by the sender. The higher the Gas price, the more likely the transaction is to be included in a block. If the Gas Limit is insufficient to pay the fee, the transaction will fail. Transactions are transparent and can be searched using transaction hashes, supporting the detection of cybercrime activities.

\noindent \textbf{Token}. Unlike native cryptocurrencies such as BTC and ETH, a token is a unit of value created and managed by smart contracts. It can serve as a cryptocurrency, voting rights, or access to services. Ethereum tokens follow the standards that determine how the tokens work and interact with other applications. According to the standards, these tokens can be categorized into fungible and non-fungible tokens (NFT). Most fungible token contracts follow the ERC-20~\cite{erc20} standard, such as Tether (USDT) and USD Coin (USDC). Each ERC-20 token is identical and can be exchanged on a one-to-one basis. Most NFT contracts follow ERC-721~\cite{erc721} and ERC-1155~\cite{erc1155} standards. Each ERC-721 token is unique and distinct, which can be distinguished from other NFTs within the same contract by a unique token ID. Therefore, each token can be identified by a contract address or the contract address with a token ID. Token transfer reflects the behavior of addresses, which is ignored by existing datasets.

\begin{figure*}[!tb]

\centering
\includegraphics[scale=0.95]{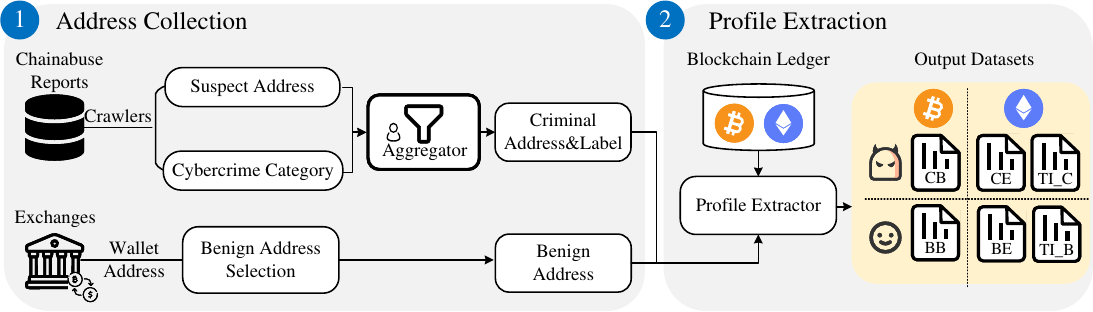}
\caption{Workflow of address collection and profile extraction process. CB and BB are criminal and benign addresses on Bitcoin. CE and BE are criminal and benign addresses on Ethereum. TI\_C and TI\_B are token interactions of criminal and benign Ethereum addresses.}
\label{fig:1}

\end{figure*}

\subsection{Related Work}
Given the substantial annual losses caused by rampant cybercrime campaigns within the blockchain ecosystem, researchers have actively explored methods to enhance the security of cryptocurrency trading. Their attentions mainly focus on the two most popular blockchain systems: Bitcoin and Ethereum. 

\noindent\textbf{Bitcoin}. In the early stage, researchers utilize address clustering heuristics to expand the cluster of the disclosed criminal addresses, discovering the associated addresses that are unrevealed~\cite{Elli2013BTCPrivacy,Meiklejohn2013noname}. Since Weber et al. introduced a labeled anti-money laundering dataset called Elliptic~\cite{weber2019elliptic} in 2019, a wealth of studies have resorted to machine learning methods to detect criminal addresses.

Feature engineering-based methods rely on the effectiveness of extracted features and utilize traditional machine learning methods to detect criminal addresses. Bartoletti et al.~\cite{Bartoletti2018ponziDetect} extract features such as lifetime and utilize machine learning algorithms like random forest to detect criminal addresses of Ponzi scheme, achieving over 97\% detection accuracy. Alarab et al.~\cite{Alarab2020comparativeML} presented an analysis on the Elliptic dataset to compare the performance of supervised learning methods such as Random Forest and AdaBoost, obtaining 0.99 precision and 0.72 recall. 

Node embedding-based methods generally model the relationship in the graph structure. Oliveira et al.~\cite{Oliveira2021GuiltyWalker} perform random walks on the Bitcoin transaction graph constructed by the Elliptic dataset and compute features based on the distance to illicit transactions, obtaining 0.97 precision and 0.77 recall. Moreover, several node embedding methods are proposed to detect illicit transactions. Loa et al.~\cite{Lo2023Inspection} propose Inspection-L, a graph neural network framework for anomaly detection, and similarly, Alarab et al.~\cite{Alarab2023GCNBit} propose a graph-based LSTM with a graph convolutional network to achieve 0.92 precision and 0.71 recall.

The sparsity of labeled criminal addresses and the imbalanced data are two challenges for cybercrime detection~\cite{Elliptic++}. For instance, the Elliptic dataset contains 203,769 node transactions, where 2\%(4,545) are labeled as illicit. The scarce labeled data is inadequate to capture the full scope of cybercrime behavior, resulting in a lack of comprehensiveness, classifiability, and real-world transferability. To alleviate this problem, two datasets, Elliptic++~\cite{Elliptic++} and Elliptic2~\cite{Elliptic2} are introduced to facilitate the cryptocurrency cybercrime detection research. Elliptic++ extends the Elliptic transaction dataset to include over 822K Bitcoin addresses, while 14,266 are illicit. Elliptic2 is a graph dataset used for subgraph representation learning. However, The real-world transferability of these two datasets remains limited, as their extensions are still based on Elliptic, which has practicality limitations. Generally, our proposed Real-CATS provides a more comprehensive labeled dataset from another source, which satisfies C3R characteristics and can be leveraged to further evaluate cybercrime detection studies and contribute to the research community.

\noindent\textbf{Ethereum}. The cybercrime detection methods on Ethereum can also be categorized into feature engineering-based methods and node embedding-based methods. 

For feature engineering-based methods, Chen et al.~\cite{Chen2021ijcai} propose a cascaded feature extraction method to aggregate information from the node and its neighbors and utilize DElightGBM to detect phishing addresses, achieving over 0.8 precision and recall. 

Node embedding-based methods are prevailing for Ethereum cybercrime detection. Trans2Vec~\cite{Wu2022Trans2vec} and its variants~\cite{trans2vecVar1,trans2vecVar2} utilize a random walk algorithm to learn the graph structure with temporal and amount information. TTAGNN~\cite{Li2022TTAGN} generates node embedding by combining the LSTM encoder and Graph Attention Network (GAT). SIEGE~\cite{Li2023SIEGE} leverages a self-supervised learning method to learn the node presentation from a large-scale graph.

These Ethereum cybercrime detection methods typically collect labeled criminal addresses from Etherscan.io (\url{https://etherscan.io}) and randomly select benign addresses. However, the absence of an effective real-world dataset makes it challenging to compare the experimental performance of these detection methods. For example, the data used in the aforementioned studies is either not fully available or only providing processed transaction graphs that are difficult to use in a customizable way. These issues make it difficult for researchers without cryptocurrency-related knowledge to directly utilize the data and participate in cybercrime detection research.

\section{Methodology}~\label{method}
In this section, we provide an overview of the data collection and processing steps of Real-CATS. The workflow is displayed in Fig.~\ref{fig:1}, consisting of two parts: (i) the address collection approach, which includes crawling abuse reports and aggregating categories to obtain labeled criminal addresses, as well as extracting benign addresses from exchange wallet addresses based on the KYC (Know Your Customer) mechanism; and (ii) querying the cryptocurrency ledger and profiling each Bitcoin and Ethereum address. Consequently, we create six data files shown in the yellow-highlighted part of Fig.~\ref{fig:1} and an identifier file for indexing.

\subsection{Address Collection}
To ensure the comprehensive distribution, we collect addresses from real-world abuse reports and the exchange customers. The address collection process can be divided into: (1) criminal address \& label collection and (2) benign address selection. 

For criminal address \& label collection, we first build website crawlers to obtain raw reports information from Chainabuse.com (\url{https://www.chainabuse.com}), encompassing reported suspected addresses and cybercrime categories until May 30, 2024. Chainabuse provides verified scam reports on Bitcoin and Ethereum, which has been utilized as a data source in several studies~\cite{He2023Txphish,Rosenquist2024Darkside,Gomez2022watch,Gomez2023Estimate}. 

Subsequently, we manually conduct category aggregation while preserving the real-world characteristics. Since the cybercrime categories are assigned by victims with varying expertise, some addresses are casually labeled. These reporters can submit custom scam types, such as "Other: Disguise as my friend", which in fact falls under "Impersonation Scam". 

We employ two ways to aggregate these custom categories. The first is using the information provided by users. For instance, we integrate various expressions into a single category (e.g., blackmail, mail scam, spam, and misspellings like "blamckail" are all classified as "Blackmail Scam"). The second is categorizing these scams by identifying certain keywords. For example, scams involving keywords like "2x," "3x," or "multiplier", which suggest multiple returns on investment, are classified as "Investment Scam". It is a broad category that includes Ponzi Schemes~\cite{Chen2018ponzi}, Rug Pulls~\cite{huang2023deepdivenftrug}, Airdrop Scams~\cite{Sayak2024airdrop} and Giveaway Scams~\cite{Li2023DoubleAN}. According to Li et al.~\cite{Li2023DoubleAN}, doubling returns fall under "Giveaway Scam". However, report information is insufficient to identify whether the scam involves giveaway behavior, so we categorize it under the broader "Investment Scam" category. Reports mentioning celebrities like "Elon Musk" are typically classified as "Impersonation Scam". Those containing social media like YouTube or Twitter are categorized as "Social Media Scam", involving scams posted on social platforms. Reports lacking detail such as "Other: BTC", "ripoff", "Defraud", and "Estorsione" are all classified under the "Other" category. We remove non-scam reports such as "Other: Sent by mistake".

For benign address selection, our method is based on the assumption that popular centralized exchanges implement strict KYC measures to combat money laundering. Therefore, selecting benign addresses from exchange customers is more reliable than random selection and more comprehensive than collecting licit addresses of service providers. We use several exchange hot wallets as seeds from Wallet Explorer (\url{https://www.walletexplorer.com}) and Etherscan.io (\url{https://etherscan.io}) to perform forward selection. For each transaction of the wallet address, we first determine whether it is an incoming or outgoing transaction. If it is an outgoing transaction, the next-hop addresses that are different from the sender address are considered benign. Note that on Ethereum, the exchange can transfer tokens through smart contracts, where the real next-hop is the token receive address rather than the contract address. In this context, we resolve the input data to determine whether a transaction is a token transfer. If the input data starts with "0xa9059cbb" (indicating \textit{transfer} function), we further resolve the input to extract the real next-hop address.

Analyzing incoming transactions is more complex because exchanges usually create a deposit address (DA) for each user to receive transfers, i.e., the previous-hop of a wallet address might still belong to the exchange rather than to a regular user. Different exchanges employ various DA mechanisms, making it challenging to collect benign addresses. Therefore, we only collect benign addresses from the outputs of exchange wallets. Besides, we delete all addresses labeled as benign and criminal simultaneously to ensure the effectiveness of benign addresses.

The address collection method we propose allows for the acquisition of more diverse address features compared to heuristic approaches, encompassing both benign user addresses and service provider addresses. Despite previous studies~\cite{Chen2021ijcai,Li2022TTAGN} filtering out addresses with too many or too few transactions, we retain all addresses, as they contribute to enhancing diversity. The diversity of normal addresses enhances the robustness of the detector and mitigates overfitting, aiding in the identification of a wider variety of addresses in real-world deployments. For example, it helps distinguish phishing criminals from ordinary traders, as well as differentiate between illegal markets and legitimate service providers.

Generally, Real-CATS comprises various criminal and benign addresses on Bitcoin and Ethereum. Criminal addresses are derived from real abuse reports submitted by victims, and include 90,612 Bitcoin addresses and 12,591 Ethereum addresses with labels, recorded as "CB" and "CE". For benign addresses, a set of 90,176 Bitcoin addresses and 16,020 Ethereum addresses is collected from the output of the exchange hot wallet, recorded as "BB" and "BE". Each address has a profile including 32 and 52 features to describe Bitcoin and Ethereum addresses, respectively. Ethereum address profile includes transaction features and token interactions. Token interactions are Ethereum address's interactions with ERC-20, ERC-721, and ERC-1155 tokens, recorded as "TI\_C" for criminal and "TI\_B" for benign. With token interactions, researchers can incorporate token interaction behavior into Ethereum address analysis. Moreover, we collect a supplementary dataset, Sup-CATS to test the comprehensiveness of our Real-CATS, including 3,147 criminal and 11,058 benign addresses on Ethereum, which simulates a real-world deployment scenario for detecting criminal addresses within a single day. Detail information of Sup-CATS is given in Section~\ref{trans}.

\begin{table}[!t]
\centering
\scriptsize
\resizebox{0.49\textwidth}{!}{
\begin{tabular}{p{0.21\textwidth}p{0.28\textwidth}}
\toprule
\textbf{Feature} & \textbf{Description} \\
\midrule
Balance & Remaining balance (Sat) of the address\\
Total\_received\_BTC & Total received bitcoins (Sat) of the address\\

Total\_sent\_BTC & Total sent bitcoins (Sat) of the address\\
Total\_received\_USD & The value of Bitcoin received in USD\\
Total\_sent\_USD & The value of Bitcoin sent in USD\\
Transaction\_fee & Total transaction fee spent (Sat)\\
Transaction\_fee\_variance & Variance of transaction fee (Sat)\\
Transaction\_number & Total number of transactions\\
Payment\_transactions & Total number of outgoing transactions\\
Receipt\_transactions & Total number of incoming transactions\\
First\_time & Timestamp of the first transaction\\
Last\_time & Timestamp of the last transaction\\
Lifetime & Time gap (seconds) from first to last transaction\\
Activity\_w & Number of days with at least one withdrawal TX\\
Activity\_d & Number of days with at least one deposit TX\\
Activity\_time & Number of days with activities\\
Max\_sent\_amount & The largest sent amount (Sat) for the address\\
Min\_sent\_amount & The smallest sent amount (Sat) for the address\\
Max\_received\_amount & The largest received amount (Sat) for the address\\
Min\_received\_amount & The smallest received amount (Sat) for the address \\
Max\_sent\_transaction\_id & Transaction hash with the largest sent amount\\
Min\_sent\_transaction\_id& Transaction hash with the smallest sent amount\\
Max\_received\_transaction\_id& Transaction hash with the largest received amount\\
Min\_received\_transaction\_id& Transaction hash with the smallest received amount\\
Received\_variance\_BTC & Variance of received amount (Sat)\\
Sent\_variance\_BTC & Variance of sent amount (Sat)\\
Received\_variance\_USD & Variance of received amount (USD)\\
Sent\_variance\_USD & Variance of sent amount (USD)\\
\bottomrule
\end{tabular}
}
\caption{\label{tab:feature} Basic profile features extracted per address on Bitcoin. Basic features of Ethereum align with Bitcoin, requiring only a change in currency units from Sat to Wei. One ETH is equivalent to $10^{18}$ Wei.}
\end{table}

\subsection{Profile Extraction} \label{3.2}

To provide the customizable data, we extract address features from transaction transaction records and release both. The profile contains 32 and 52 features for Bitcoin and Ethereum addresses, respectively. In this section, we describe all features of the profile.

\noindent \textbf{Bitcoin}. We run a full node and download the cryptocurrency ledger to obtain transaction information. We iterate through the transactions in the ledger. If a transaction involves any address in our address set, it will be recorded for further processing. Finally, we collect 39,752,469 transactions. Table~\ref{tab:feature} lists all basic transaction features we have extracted with their descriptions. Some features have been proposed by Gomez et al.~\cite{Gomez2022watch} to detect exchange addresses.

\begin{table}[!t]
\centering
\scriptsize
\resizebox{0.49\textwidth}{!}{
\begin{tabular}{>{\centering\arraybackslash}p{0.06\textwidth} p{0.10\textwidth} p{0.33\textwidth}}
\toprule
\textbf{Blockchain} & \textbf{Feature} & \textbf{Description} \\ \midrule
\multirow{4}{*}{Bitcoin} & Total\_input\_slots & Total input slots in which the address appear \\  
        & Total\_output\_slots & Total output slots in which the address appears \\  
        & Sent\_counters & Number of addresses used as input slots for interaction \\  
        & Received\_counters & Number of addresses used as output slots for interaction \\  
\midrule
\multirow{9}{*}{Ethereum} & Create\_contract & Number of contract creation \\  
        & Error\_count & Number of error transactions \\  
        & Total\_gas\_used & Total amount of used gas \\  
        & Total\_gas\_limit & Total amount of gas set in the transaction \\  
        & Total\_gasP\_set & Total amount of gasPrice set in the transaction \\  
        & Gas\_limit\_Variance & Variance of the set gas \\  
        & GasP\_set\_Variance & Variance of the set gasPrice \\  
        & Gas\_weight\_block & Weight of average transaction cost per block \\  
        & ERC Interactions & Including 16 features to measure the specific interaction. \\  
\bottomrule
\end{tabular}
}

\caption{\label{tab:feature2} Customized profile features extracted according to the characteristics of the specific blockchain and cryptocurrency. The complete profile consists of basic features in Table~\ref{tab:feature} and customized features.}
\end{table}

We utilize the exchange rate data from \href{https://www.investing.com}{Investing.com}. The features such as $Total\_received\_USD$ are calculated with daily exchange rate to obtain more accurate results:

\begin{equation}
    Total\_received\_USD = \sum_{First\_time\leq d \leq Last\_time} Rec_d*ER_d
\end{equation}

where $Rec_d$ and $ER_d$ indicate the received amount and exchange rate in day $d$. Given the fluctuations on $ER_d$, there may be a slight variance in the results. 

Considering the structure of Bitcoin transactions, Bitcoin address profiles include four customized features, as shown in Table~\ref{tab:feature2}. During the iteration, we count the number of times an address serves as an input or output address during the iteration to extract the values for $Total\_input\_slots$ and $Total\_output\_slots$. For $Sent\_counters$, if the target address is a recipient in a confirmed transaction, we increment the count for this feature by the number of senders, excluding the address itself. Conversely, $Received\_counters$ tracks the number of recipients associated with each address.

If an address is reported as criminal without any transactions (it is possible, e.g., if the reporter received a scam email but did not fall victim to it), then the $Transaction\_number$ and other related features will be 0, and features like $First\_time$ will be empty.

\noindent\textbf{Ethereum}. We launch an Ethereum client to obtain all 23,436,690 related transactions. The Ethereum transaction data is resolved into five sections: internal, normal transactions, ERC-20, ERC-721, and ERC-1155.

For internal and normal transactions, we first distinguish between Externally Owned Accounts (EOA) and contract addresses. Subsequently, we extract basic and customized features from the internal and normal transaction data. Note that we exclude internal transactions related to contract creation. Ethereum basic features in this part are similar to the one used for Bitcoin addresses, only the currency units are changed to Wei. The $Transaction\_fee$ and $Transaction\_fee\_varience$ are extracted by calculating the sum and variance of the product of gas used and gas price for each transaction. However, the transaction fee-related features for contract addresses are all zero, as they are not required to pay fees for transactions.

\begin{table}[!tb]
\centering

\scriptsize
\begin{tabular}{llll}
\toprule
\textbf{Address} & \textbf{ERC-20} & \textbf{ERC-721} & \textbf{ERC-1155} \\
\midrule
0x2c9e6... & 607641:1T2, 401070:1T1& - & 18262:0T1\\

0x407ce... & 528854:0T1& 1726:1T1 & 32675:0T1\\

0xa98c0... & 463097:1T3 & 411025:1T1, 36023:1T1 & - \\
\bottomrule
\end{tabular}

\caption{\label{tab:inter} Token interaction behavior table (example). "607641" points to the token contract address saved in row 607,641 in "Identifier.tsv". "1T2" indicates that the address served as the input address for the contract once and as the output address twice. }

\end{table}

\begin{table}[!t]
\centering
\scriptsize
\begin{tabular}{ll}
\toprule
\textbf{Index}&\textbf{Identifier} \\
\midrule
1&0x90685e300a4c4532efcefe91202dfe1dfd572f47\\
2&0x9f9b2b8e268d06dc67f0f76627654b80e219e1d6\_7764\\
3&0xcf7c0ec7c2cb5bb7392be55953ed14064ff2a4c2\_2737\\
$\cdots$&$\cdots$\\
910,864 & 0xb817d88dbfa9039c5c15b31712c415e0e0d24904\\
\bottomrule
\end{tabular}
\caption{\label{tab:id} The structure of "Identifier.tsv", including all related tokens. An identifier is a token contract address and, if the contract is ERC721 or ERC1155, append the token ID after the underscore. A row index in Table~\ref{tab:inter} refers to an entry. }
\end{table}

Since Ethereum has a diverse ecosystem such as contracts and tokens, we extract additional profile features to illustrate the traces left by a cybercriminal. Table~\ref{tab:feature2} shows these customized features. When an Ethereum user initiates a transaction, they can decide the maximum amount of gas to spend and the price per unit of gas. Therefore, we extract gas-related information to reflect a user’s urgency in completing the transaction. Furthermore, we use a group of 16 features to show the specific interaction frequency. We define a 4-tuple $(action, with\_contract, with\_data, transaction\_type)$ to describe 16 types of interaction. The \textit{action} denotes whether an address is sender or receiver, \textit{with\_contract} signifies whether the transaction interacts with a contract, \textit{with\_data} indicates whether the transaction has input data, and \textit{transaction\_type} denotes whether the interaction occurred in an internal or normal transaction. For example, the feature $From\_contract\_wod\_internal$ counts the number of internal transactions with the smart contract, where this address acted as the sender and no data input was involved.

For ERC-20, ERC-721, and ERC-1155, we extract the token interaction behavior. Token interaction behavior is another important part of Ethereum user profiles, which is overlooked by existing datasets. Initially, we retrieve all Token-related transactions of the collected addresses, if applicable. Subsequently, we identify and record the token contract addresses that interact with the collected addresses (and record token ID if the token follows ERC-721 or ERC-1155). Finally, we calculate and record the interaction behavior between the collected addresses and token contracts in "TI\_C" and "TI\_B". The data structures of two files are listed in Table~\ref{tab:inter}. We simplify the entries by saving all 910,864 token contract addresses in the file "Identifier.tsv", as shown in Table~\ref{tab:id}. Note that if all 52 features of an Ethereum address are zero or null, it could indicate either a failed criminal address or an address involved only in token transactions.

\section{Data Analysis}

In this section, we validate C3R characteristics and demonstrate the functionalities of Real-CATS. Since the comprehensiveness of Real-CATS is mainly provided by the diversity and richness of the victims, we merge the analysis of comprehensiveness with classifiability. All address information can be searched on the blockchain browser.

\subsection{Comprehensiveness and Classifiability} \label{sec4.1}

\begin{figure*}[!tb]
	\centering 
	\subfloat[Receive-Sent of Criminal Address]{
		\includegraphics[width=0.24\linewidth]{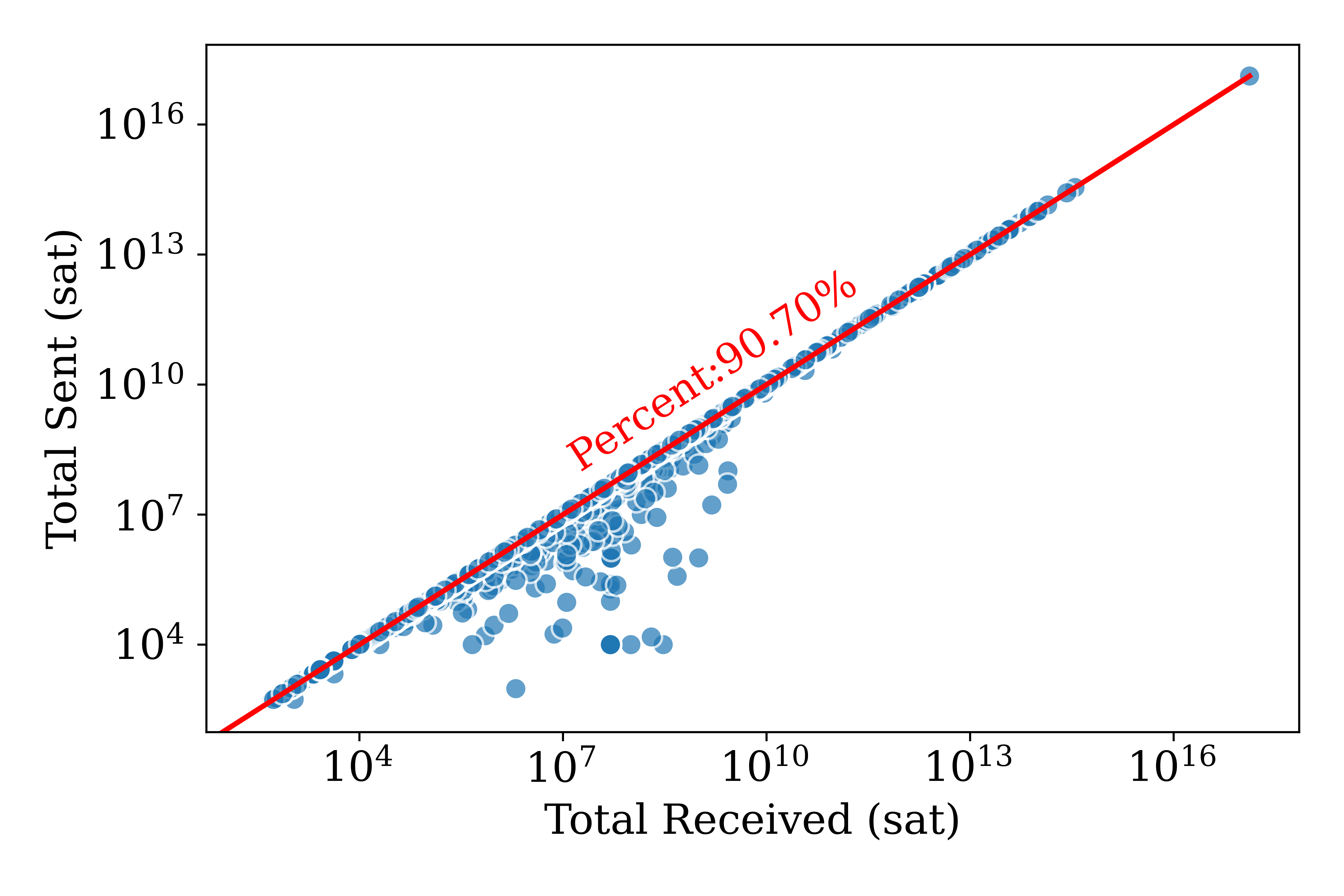}\label{fig2a}
}
	\subfloat[Receive-Sent of Benign Address]{
		\includegraphics[width=0.24\linewidth]{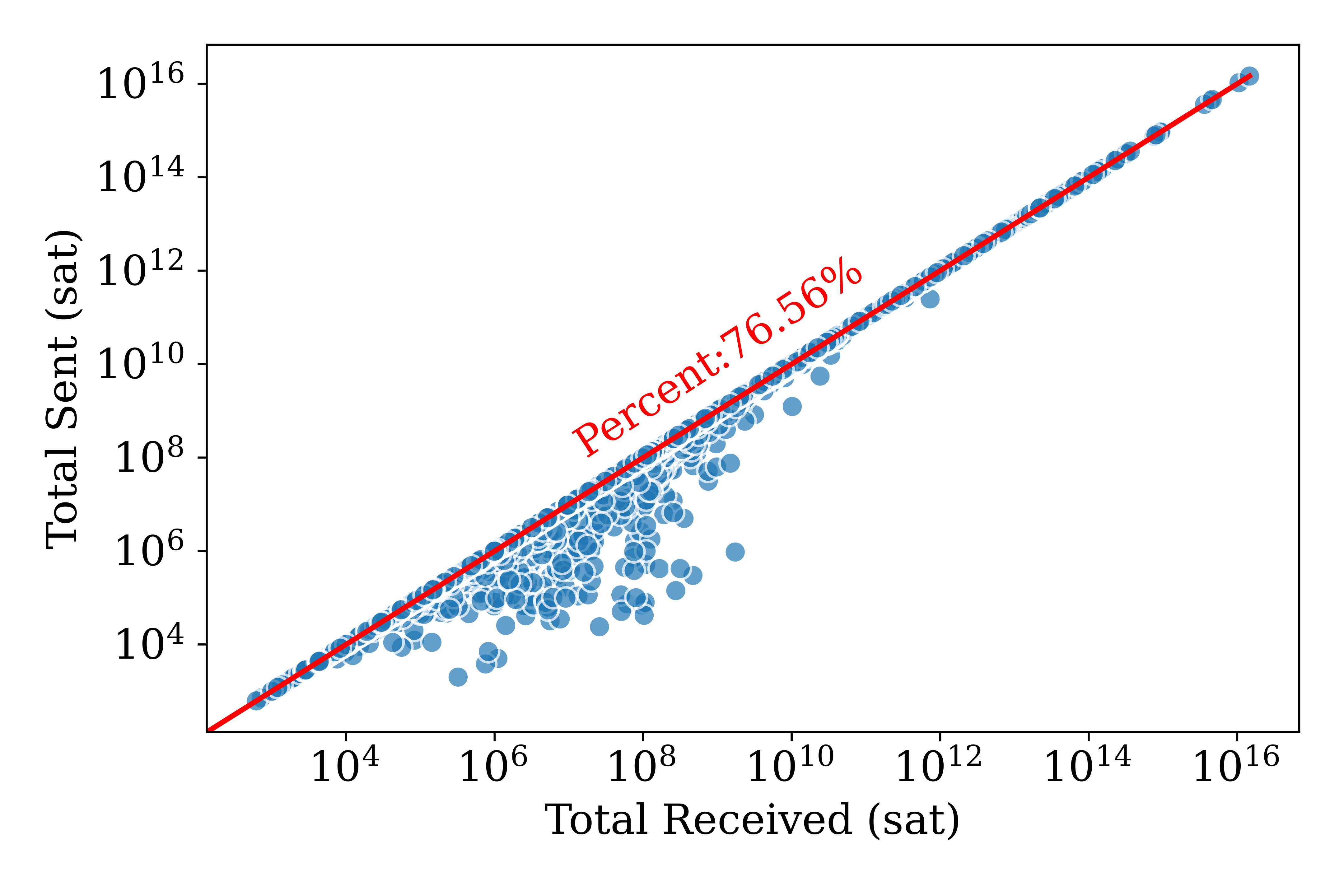}\label{fig2b}
}
        \subfloat[$Lifetime$ of Criminal Address]{
		\includegraphics[width=0.24\linewidth]{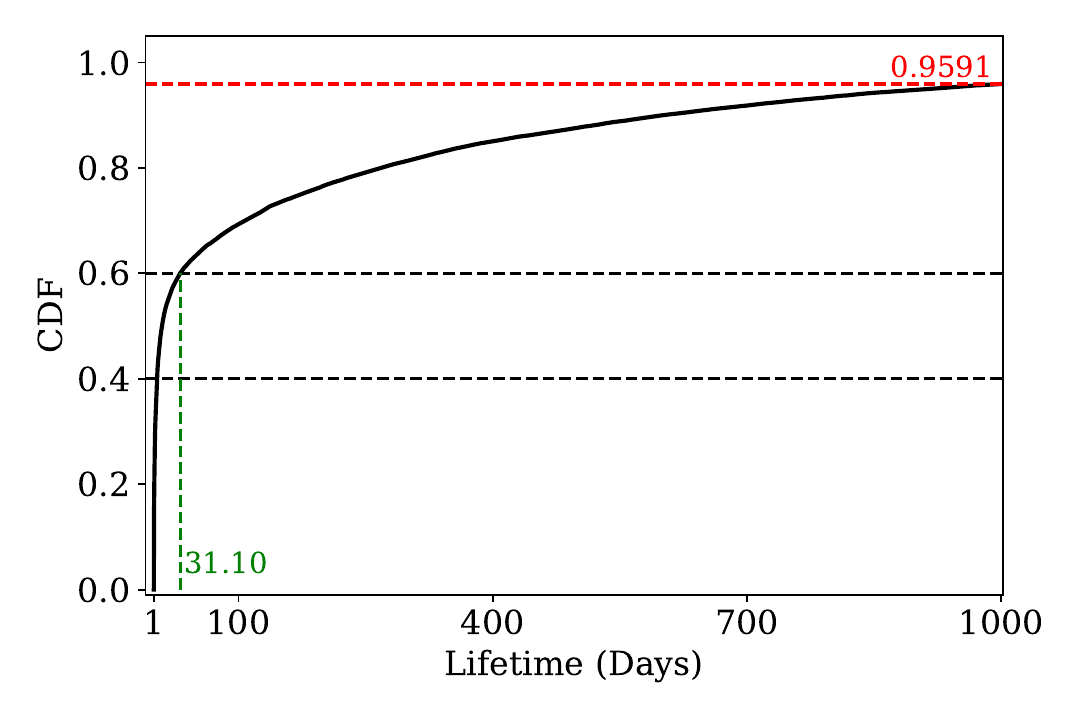}\label{fig2c}
}
	\subfloat[$Lifetime$ of Benign Address]{
		\includegraphics[width=0.24\linewidth]{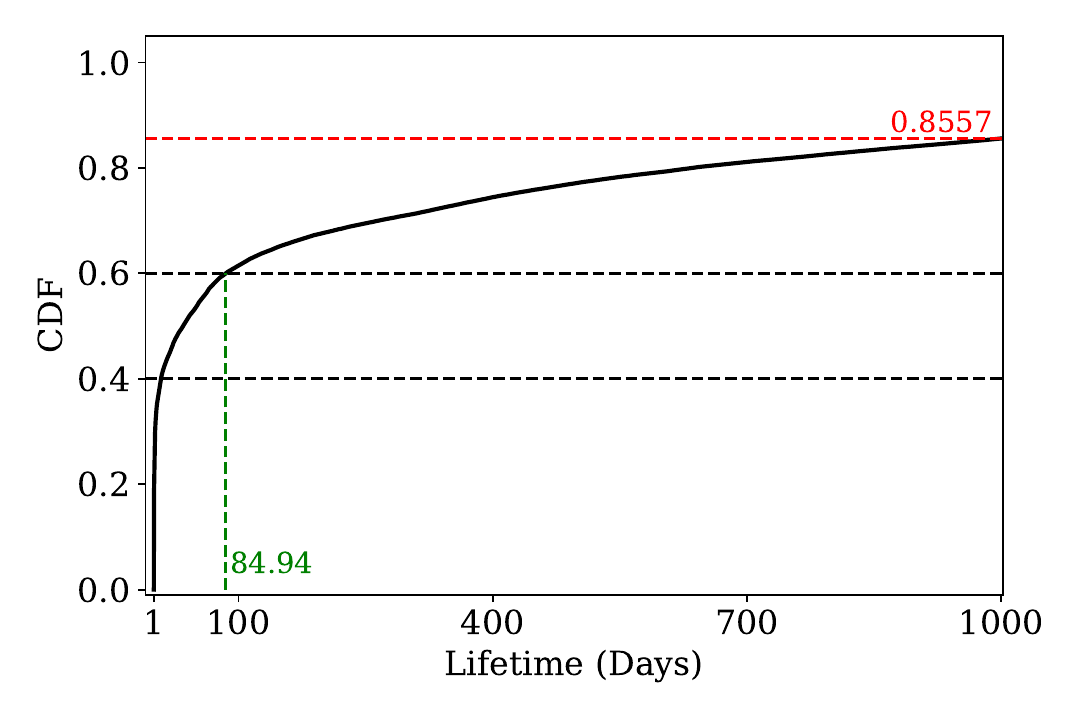}\label{fig2d}
}
    \caption{Scatterplot of received and sent Sat and the CDF of $Lifetime$ for criminal and benign addresses on \textbf{Bitcoin}. Red line indicates received number equals sent number. Red text is the percentage of zero balance addresses. The received vs. sent distribution indicates that criminal addresses are more likely to transfer all funds, and benign addresses tend to last longer than criminal addresses.}

    \label{fig:2}
\end{figure*}

In this section, we show that Real-CATS, which contains criminal and benign addresses that exhibit behavioral differences and align with existing conclusions, satisfies Comprehensiveness and Classifiability characteristics and is effective for cybercrime detection tasks. We analyze two patterns both in Real-CATS and in the public dataset~\cite{weber2019elliptic, Li2020illicit} and demonstrate that our dataset encompasses a broader data distribution, thus avoiding the overly apparent patterns present in existing datasets. Furthermore, we perform classification and visualization analysis to further validate the classifiability of the transaction profiles. These tasks also show the function of effective evaluation and comparison of cybercrime detection methods.

\begin{figure}[!tb]
    \centering 
    \subfloat[Receive-Sent]{
		\includegraphics[width=0.80\linewidth]{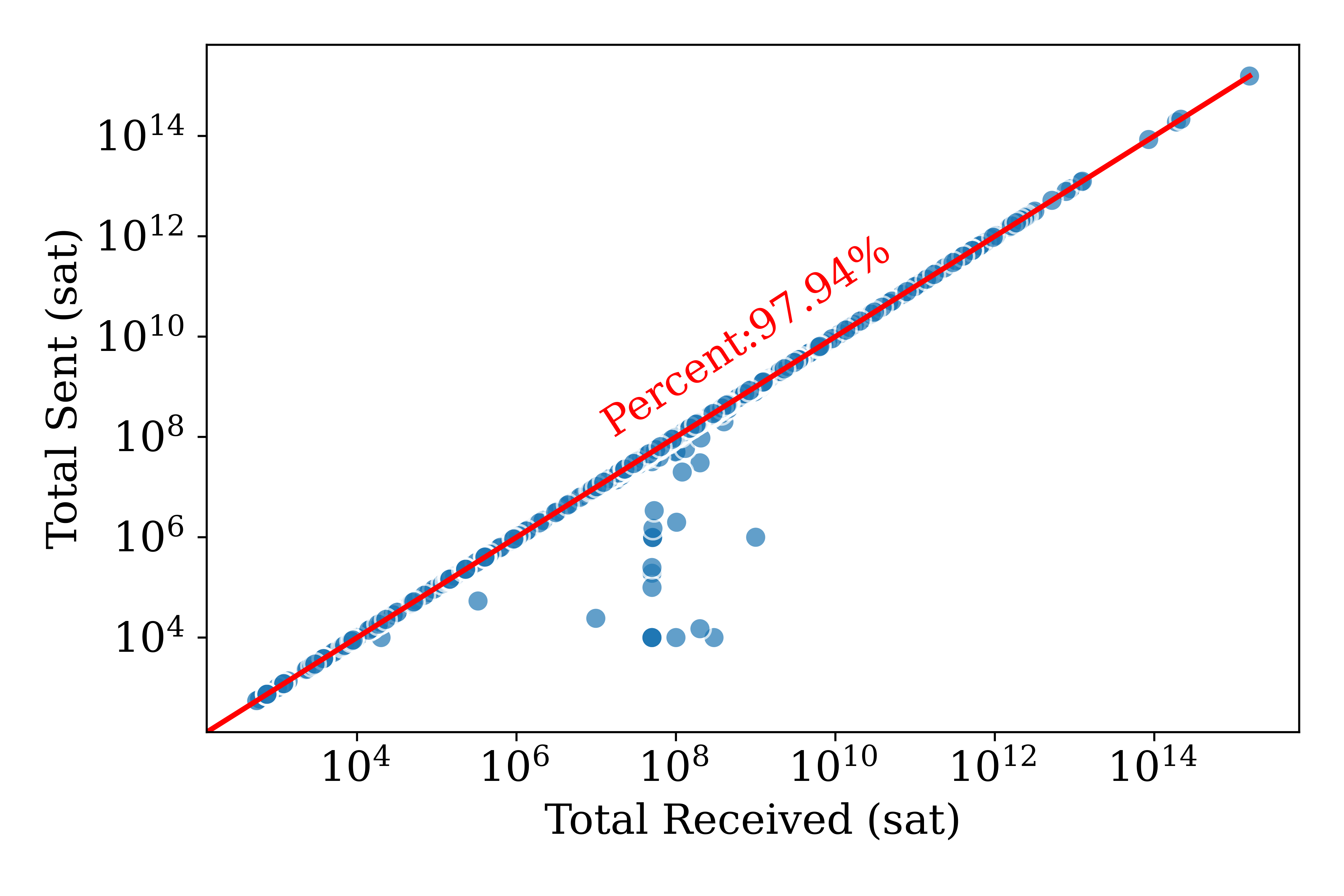}\label{other_rec}}\\
    \subfloat[$Lifetime$]{
		\includegraphics[width=0.80\linewidth]{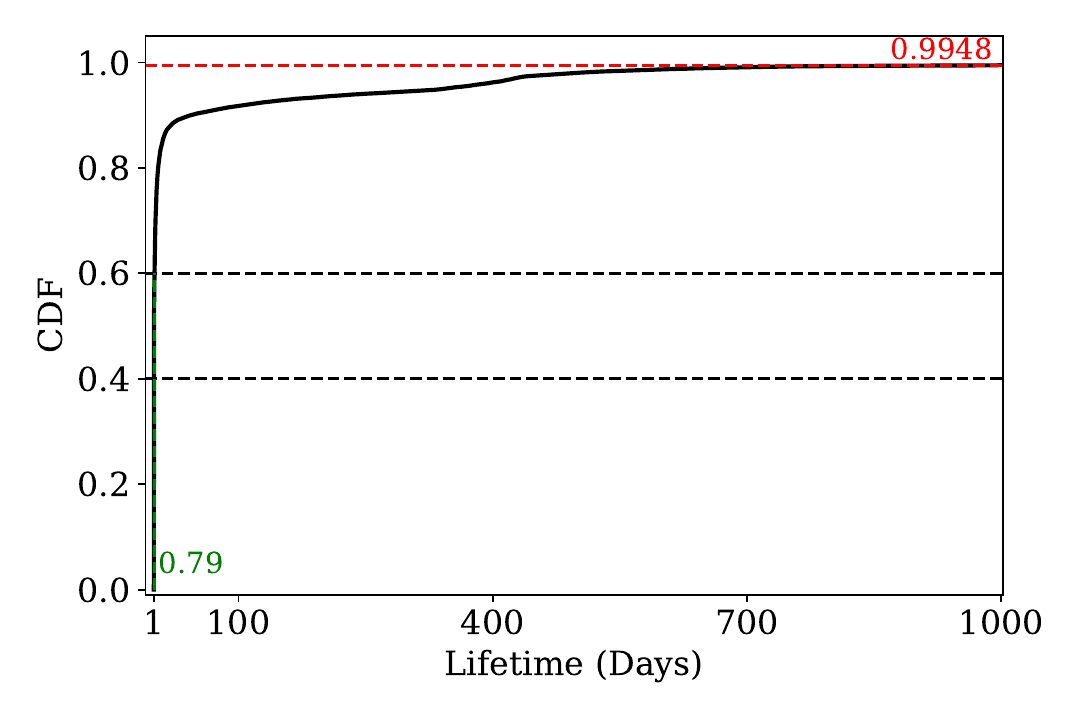}
            \label{otherlifetime}}
    \caption{Scatterplot of received and sent Sat and the CDF of $Lifetime$ for criminal addresses in \textbf{existing datasets}. Red line indicates the balance is zero where the received number equals the sent number. Red text is the percentage of zero balance addresses. Criminal addresses present overly apparent features in existing datasets.}

    \label{fig:3}
\end{figure}

\begin{figure*}[!tb]
	\centering 
	\subfloat[Receive-Sent of Criminal Address]{
		\includegraphics[width=0.24\linewidth]{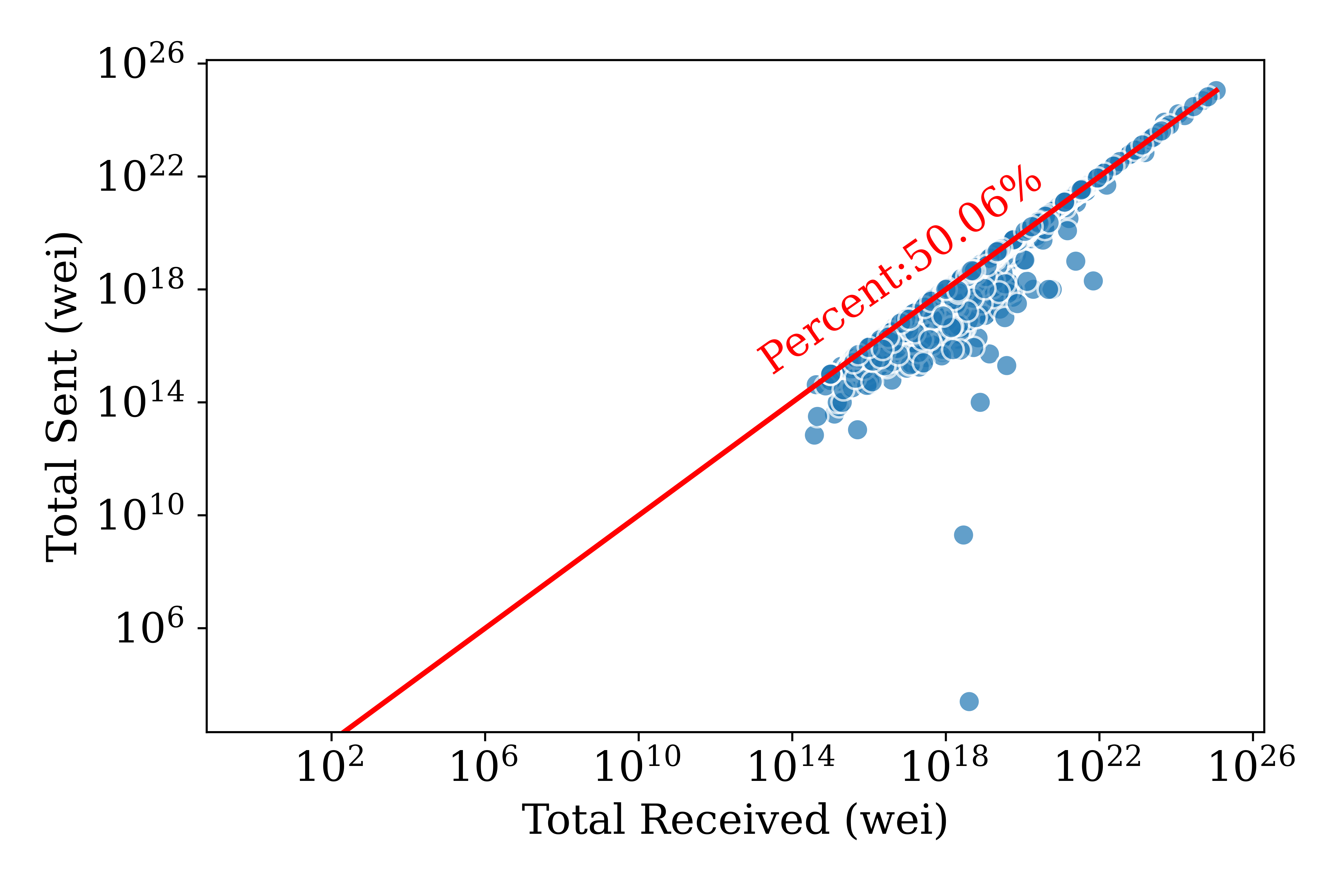}
}
	\subfloat[Receive-Sent of Benign Address]{
		\includegraphics[width=0.24\linewidth]{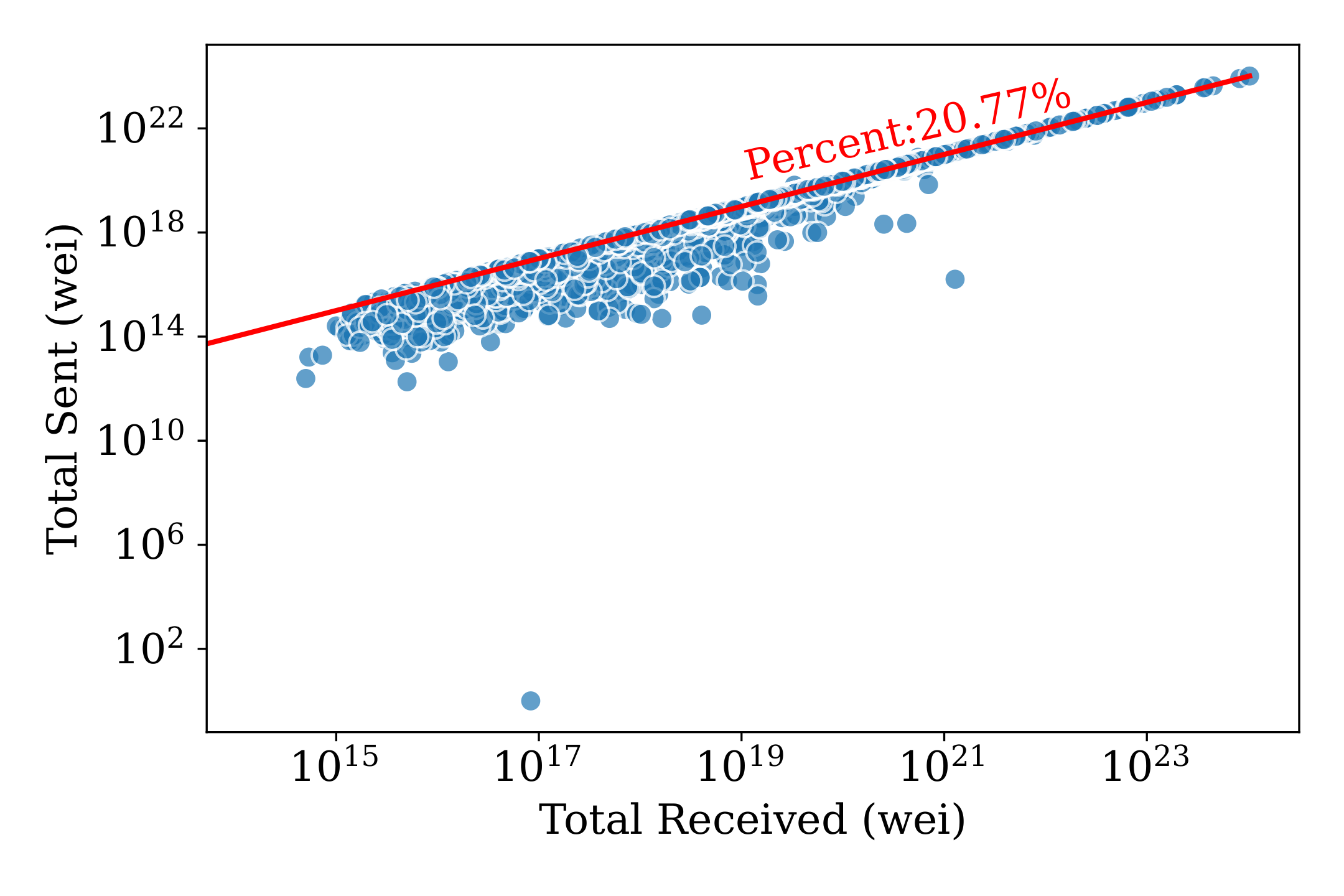}
}
    \subfloat[$Lifetime$ of Criminal Address]{
		\includegraphics[width=0.24\linewidth]{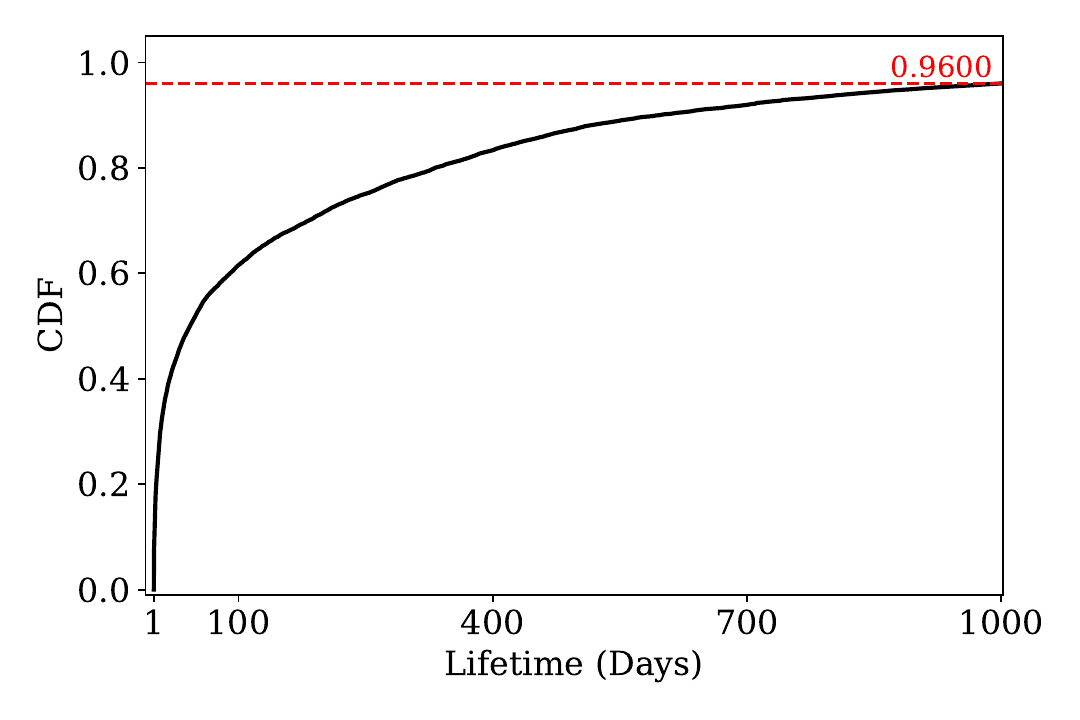}}
	\subfloat[$Lifetime$ of Benign Address]{
		\includegraphics[width=0.24\linewidth]{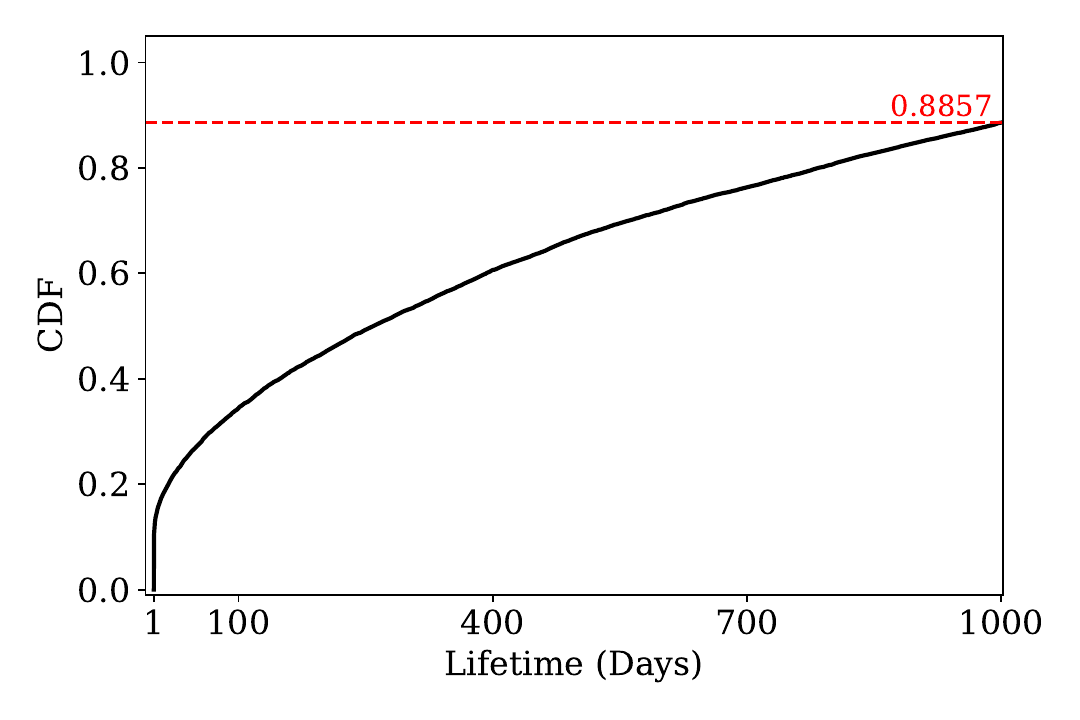}}
    \caption{Scatterplot of received and sent Wei and the CDF of $Lifetime$ for criminal and benign addresses on \textbf{Ethereum}. Red line indicates received number equals sent number. Red text is the percentage of zero balance addresses. The received vs. sent on Ethereum is evidently different. And the benign addresses tend to last longer than criminal addresses.}

    \label{fig:6}
\end{figure*}

\noindent\textbf{Patterns Analysis}. We first validate the Bitcoin data. According to Rosenquist et al.~\cite{Rosenquist2024Darkside}, nearly all reported addresses have sent as many bitcoins as they received, leaving a balance of zero. This is because the reported addresses are typically temporary addresses created by criminals, with their illicit revenues quickly sent out. We show the sent vs. received amount of addresses in Fig.~\ref{fig2a} and~\ref{fig2b}. To ensure fairness in comparison, we only consider the address with $Transaction\_number > 0$. 

Fig.~\ref{fig2a} indicates that in Real-CATS, 90.70\% of criminal addresses on Bitcoin have a balance of 0, which aligns with the temporary nature of scam addresses. In Fig.~\ref{fig2b}, temporary wallets in of benign addresses on Bitcoin only account for 76.56\%. This result demonstrates the discrepancy between criminal and benign addresses in terms of the balance pattern, showing the classifiability of Real-CATS.

The short lifespan is also an important feature of phishing addresses~\cite {He2023Txphish,Liu2024Fish}. For other criminal addresses, it is also necessary to expedite the transfer of criminal revenue. We illustrate the distribution of $Lifetime$ feature to validate the Real-CATS by the Short-Life pattern. Fig.~\ref{fig2c} and~\ref{fig2d} show the Cumulative Distribution Function (CDF) of $Lifetime$. When the $Lifetime$ is less than 5 days, the CDF curves for criminal addresses and benign addresses are initially rising at a similar rate. However, the former increases more rapidly, with 60\% of criminal addresses lasting less than 31.10 days, compared to 60\% of benign addresses lasting less than 84.94 days. When larger than 100 days, the curve of benign addresses obviously rises more gently than criminal addresses. Additionally, only 4.09\% of criminal addresses last more than 1,000 days, while 14.43\% of benign addresses exceed 1,000 days, which is three times that of criminal addresses. These curves mean that benign addresses on Bitcoin tend to have a longer lifespan, demonstrating the characteristics of criminal addresses and further showing the classifiability.

We further compare these features with the Elliptic~\cite{weber2019elliptic} dataset and the illicit address dataset~\cite{Li2020illicit}. We merge these two datasets and remove duplicates to have 21,346 criminal addresses, then plot the received vs. sent and the CDF in Fig.~\ref{fig:3}. Almost all addresses have a zero balance and a short lifetime. Compared to Real-CATS, existing datasets overly align with the two patterns mentioned above. These results indicate that existing datasets are too pure to cover the distribution of real-world behavior of criminal addresses. For example, the $Lifetime$ can effectively distinguish the criminal address according to Fig.~\ref{otherlifetime}, which is not realistic. In contrast, Real-CATS does not present overly apparent patterns, which indicates that the dataset encompasses a rich variety of criminal and benign behaviors that exhibit the Comprehensiveness characteristic.

\begin{figure*}[!tb]
	\centering 
	\subfloat[Bitcoin]{
		\includegraphics[width=0.44\linewidth]{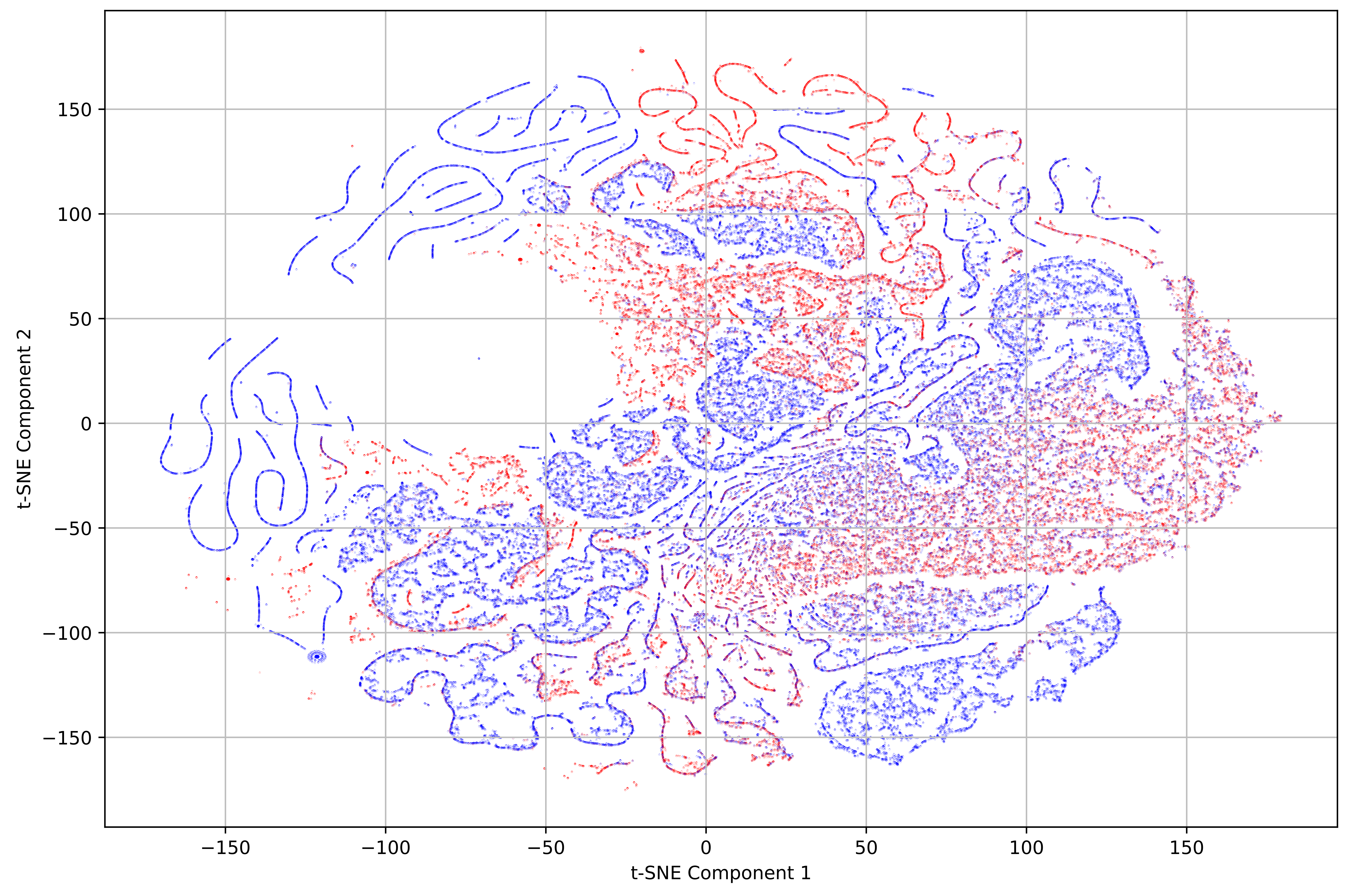}}
    \hfil \hspace{-5pt}
	\subfloat[Ethereum]{
		\includegraphics[width=0.44\linewidth]{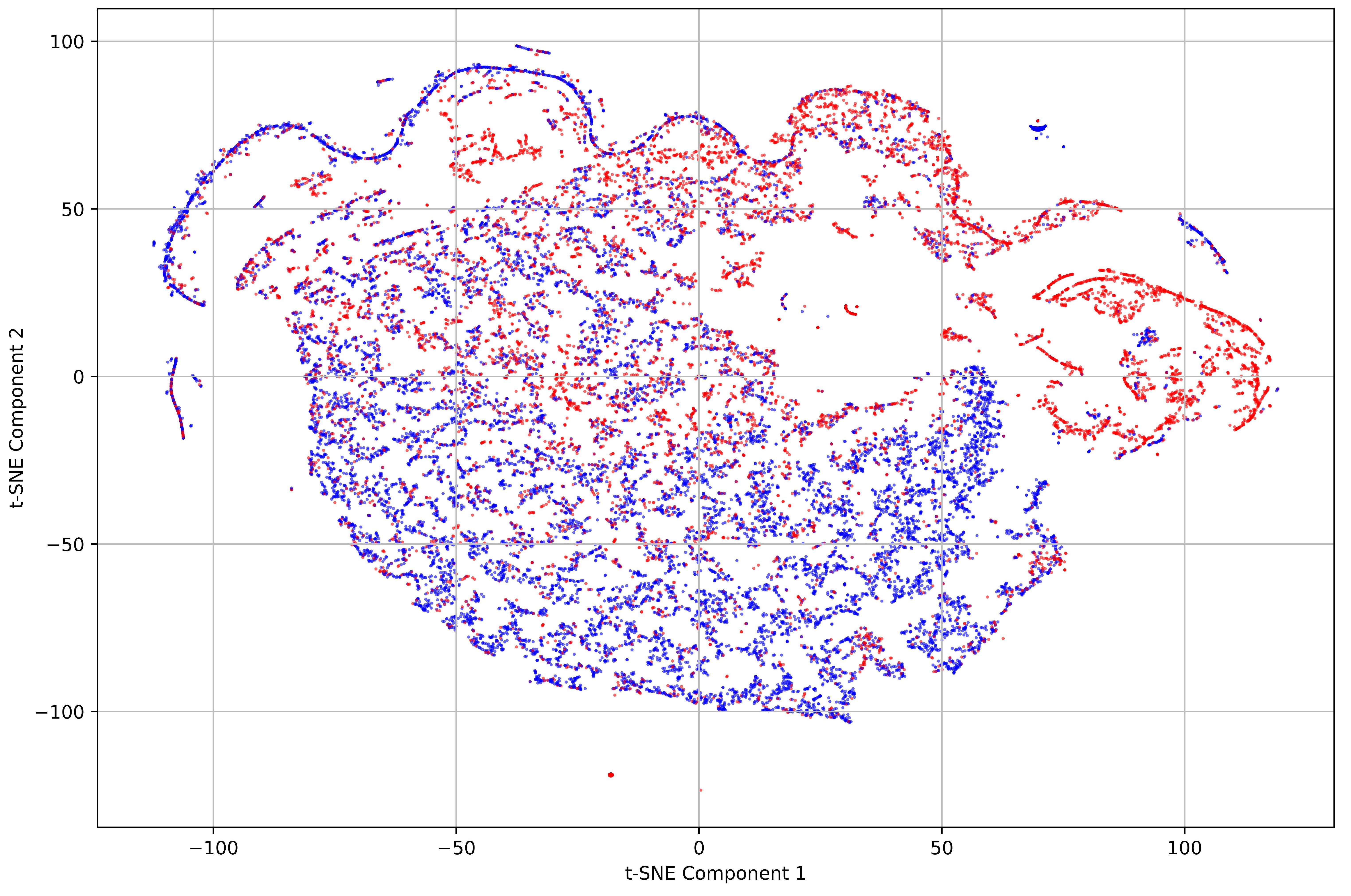}}
    \caption{t-SNE visualization of address profiles on both blockchains, showing the effectiveness of our profiles. Red points indicate criminal addresses. Blue points are benign addresses.}

    \label{fig:8}
\end{figure*}

The behavior patterns of criminal addresses on Ethereum exhibit certain similarities to those on Bitcoin, thus we still rely on the two observed behavior patterns from Bitcoin. The balance of Ethereum addresses is quite different from Bitcoin due to a more sophisticated ecosystem. In Fig.~\ref{fig:6}, only 50.06\% criminal addresses and 20.77\% benign addresses have zero balance. Although significantly lower than the proportion of zero-balance of Bitcoin addresses, the discrepancy between criminal and benign addresses also validates the effectiveness of Real-CATS. Furthermore, criminal addresses with $Lifetime$ in 100 days account for nearly 60\%, while benign addresses only for 34.09\%. The curve of criminal addresses rises more rapidly than the benign ones, meeting the Short-Life pattern. 

Despite the distribution of two features aligning with the patterns drawn from existing research, our Real-CATS dataset presents a greater challenge because the addresses are sourced from real-world abuse reports, which encompass more comprehensive behaviors.

\begin{table*}
    \centering
    \setlength{\tabcolsep}{8pt}
    \begin{tabular}{lcccccccc}
        \toprule
        \multirow{2}{*}{\textbf{Method}} & \multicolumn{4}{c}{\textbf{Ethereum}} & \multicolumn{4}{c}{\textbf{Bitcoin}} \\
        \cmidrule(lr){2-5} \cmidrule(lr){6-9}
         & \textbf{Precision} & \textbf{Recall} & \textbf{F1-score} & \textbf{Accuracy} & \textbf{Precision} & \textbf{Recall} & \textbf{F1-score} & \textbf{Accuracy}\\
        \midrule

        SVM& 0.8710 & 0.3522& 0.5015 & 0.7116 &  0.6890& 0.1750 & 0.2800 & 0.7142 \\
        $k$-NN & 0.7854  & 0.6508& 0.7117 & 0.7828 & 0.7672& 0.7717 & 0.7689 & 0.7717\\
        Decision Tree & 0.7825 & 0.7834& 0.7829 & 0.8210 & 0.6985& 0.6997 & 0.6991 & 0.8096 \\
        Random Forest & 0.9085 & 0.8237& 0.8640 & 0.7116 & \textbf{0.8294}& 0.7234 & 0.7726 & 0.8654\\
        LightGBM & \textbf{0.9187} & \textbf{0.8520}& \textbf{0.8840} & \textbf{0.9080} & 0.8130& \textbf{0.7712} & \textbf{0.7916} & \textbf{0.8716}\\
        \bottomrule
    \end{tabular}
 
    \caption{The performance comparison between five machine learning models to show the effectiveness of our profiles.}

    \label{tab:ml}
\end{table*}

\noindent\textbf{Classifiable Profiles}\label{4.1}. We utilize the t-SNE~\cite{tsne} algorithm to present the visualization of address profiles. We select and standardize the numerical features for visualization. In Fig.~\ref{fig:8}, the result shows that most of the positive and negative samples exhibit a clear distinction. 

We further leverage these numerical features and train machine learning models to evaluate the classification performance. Metrics including Precision, Recall, F1-score, and Accuracy are chosen to evaluate the detection performance. Note that the former three metrics are only for criminal addresses, which better reflect the real performance in a class-imbalanced scenario. We use the implication in sklearn, then select the RBF kernel function for SVM, set the number of neighbors to 5 for $k$-NN, and set the number of estimators to 100 for Random Forest. The dataset is filtered by removing addresses without transactions and then split into training and testing sets with a 7:3 ratio. We keep the class imbalance in the training set to better fit the imbalance in the real world, i.e., the criminals are significantly less than benign users. 

The mean metrics after a 5-fold cross validation are shown in Table~\ref{tab:ml}. Note that absolute timestamp features, such as $First\_time$, are excluded from the training process as they can harm the detector's generalization to future data. For example, the detector may learn that an address being active during a specific historical time period is a key feature for classification, which becomes irrelevant when detecting addresses in the future.

LightGBM algorithm achieves the best performance on the Ethereum dataset, with over 90\% accuracy. It also achieves the highest Recall, F1-score, and Accuracy on the Bitcoin dataset. This result indicates that the transaction profiles provided by our dataset offer sufficient information, enabling good distinguishability between the two types of addresses. These transaction profiles not only satisfy classifiability but also lay the foundation for customizability.

Overall, these demonstrations show that Real-CATS is effective for evaluating cybercrime detection methods. It provides researchers with a common benchmark for algorithms and a shared evaluation platform, allowing them to benefit from and contribute to the research.

\subsection{Customizability}

As mentioned in Section~\ref{3.2}, the customizability characteristic of Real-CATS is mainly provided by releasing both statistical transaction profiles and detailed transaction records. In this section, we indicate that the Real-CATS dataset can be extended using detailed transaction records to customize the data and foster the discovery of new features and innovative methods. 

As an example, we can customize the address-specific Real-CATS into a large-scale transaction graph. Although the transaction profiles do not provide every single transaction related to an address, we can extract this information from the detailed transaction records and construct input data with richer information. For instance, we can extract each transaction's participants, timestamps, and amounts of each address to build a weighted multidigraph.

To demonstrate the effectiveness of this customization, we randomly sample a subset of our dataset and use the 1-order neighbors to construct a transaction graph and test the performance of random walk methods. We resort to the script provided by Wu et al.~\cite{Wu2022Trans2vec} to detect the criminal addresses. This subset only contains 2,500 addresses with a ratio of 1:1 between criminal and benign addresses and contains 47,009 addresses after extension. The ratio of training and testing set is 7:3. The parameters are the same as in the original paper. 

Trans2Vec~\cite{Wu2022Trans2vec} achieves 0.8976 Precision, 0.9261 Recall, 0.9116 F1-score, and 0.9348 Accuracy. These results outperform the most effective profile-based detector LightGBM and indicate that our Real-CATS dataset is well-suited for node embedding-based methods once extended to a large transaction graph, proving the customizability of our dataset. 

However, node embedding-based methods require the construction of large-scale transaction graphs, which entail significant overhead for data processing and training. Furthermore, these transductive methods may be challenging to apply for real-time detection of new addresses~\cite{Li2023SIEGE}, so we did not test them in the real-world scenario. We have also released the transaction records of each address in Real-CATS publicly available.

\subsection{Real-world Transferability} \label{trans}

Real-CATS is a dataset derived from abuse reports. Currently, authoritative platforms like Etherscan also provide labeled criminal addresses, which are important data sources for cybercrime detection~\cite{Wu2022Trans2vec,Yang2024DynEth}. However, these studies often evaluate by simply splitting the dataset proportionally, neglecting the temporal relationships in real-world deployments, where training samples are typically slightly older than the detection targets. In this section, we introduce a supplementary dataset, named Sup-CATS, to simulate a real-world deployment scenario, where the model is deployed to scan all addresses active on a given day and detect criminal addresses. This dataset allows for the evaluation of the classifier's effectiveness in real-world environments.

To simulate a real-world deployment environment, we further collect a supplementary dataset for testing, named Sup-CATS. Sup-CATS comprises crime and benign addresses sampled from all blocks on August 12, 2024. It encompasses 3,147 criminal addresses labeled by Etherscan. We scan and sample the unlabeled addresses as benign with a 1\%  probability from all 7,161 blocks on the same day. We finally collect 11,058 benign addresses in total. Note that we only collect labeled addresses on Ethereum, as obtaining authoritative labels on Bitcoin is more challenging.

\begin{table}[]
\begin{tabular}{cccrc}
\toprule
Method               &      Prediction      & \multicolumn{2}{c}{Truth}                          & Class error                \\
\cmidrule(lr){2-2} \cmidrule(lr){3-4}  \cmidrule(lr){5-5} 
\multirow{3}{*}{\begin{tabular}[c]{@{}c@{}}\textbf{Random}\\ \textbf{Forest}\end{tabular}} &  & FP & \multicolumn{1}{c}{TP} &  \\
                      & FP         & \multicolumn{1}{r}{8,148} & 2,907                  & \multicolumn{1}{r}{26.3\%} \\
                      & TP         & \multicolumn{1}{r}{790}   & 2,357                  & \multicolumn{1}{r}{\textbf{25.1\%}} \\
\midrule
Method               &       Prediction     & \multicolumn{2}{c}{Truth}                          & Class error                \\
 \cmidrule(lr){2-2} \cmidrule(lr){3-4}  \cmidrule(lr){5-5} 
\multirow{3}{*}{\textbf{LGBM}} &  & FP                        & \multicolumn{1}{c}{TP} &                            \\
                      & FP         & \multicolumn{1}{r}{8,379} & 2,676                  & \multicolumn{1}{r}{\textbf{24.2\%}} \\
                      & TP         & \multicolumn{1}{r}{2,834} & 313                    & \multicolumn{1}{r}{90.1\%} \\
\bottomrule 
\end{tabular}%

\caption{Confusion matrix of classifying the addresses in Sup-CATS with Random Forest and LGBM trained on Real-CATS.}
\label{tab:cm}
\end{table}

Subsequently, we split the Real-CATS dataset and train Random Forest and LGBM classifier with the same setting in Section~\ref{4.1}, as they perform better when presenting the effectiveness of our profile. Results in Table~\ref{tab:cm} show the capability of detecting unseen criminal addresses. However, LGBM performs significantly worse for detecting unseen criminal addresses in Sup-CATS, although it achieves great performance on the Real-CATS dataset. It merely achieves 9.9\% precision, indicating an awful detection capability. In contrast, Random Forest is more robust in this task, detecting 74.9\% of the criminal addresses.

The experimental results above show that relying solely on the transaction profiles provided in our dataset can detect over 70\% criminal addresses within a single day in the real-world scenario. It demonstrates that Real-CATS allows researchers to train their algorithms on older datasets and then test them in practical environments, assessing the model's ability to transfer to real-world scenarios. Although the high false positive rate severely affects the detection performance, recent advancements in cybercrime detection lead us to believe that these false positives can be effectively mitigated by designing a more complex and systematic criminal address filtering method.

\section{Discussion}
In this section, we discuss the application scenarios of Real-CATS and Sup-CATS. 

The primary purpose of collecting these datasets is to foster efforts to combat cryptocurrency cybercrime by attracting researchers from fields lacking cryptocurrency knowledge to contribute to the research. First, Real-CATS exhibits C3R characteristics, providing a shared platform for effective evaluation and comparison of cybercrime detection research. Furthermore, these studies can leverage the Real-world Transferability of Real-CATS to extend their evaluation. In addition, we provide transaction profiles and detailed transaction records, which not only allow anti-cybercrime practitioners to utilize these features but also enable uncovering new criminal behavior patterns. Lastly, other types of anti-money laundering research, such as cybercrime revenue estimation~\cite{Gomez2023Estimate} and money flow tracking~\cite{Gomez2022watch,Wu2024assetflow}, can also use this dataset to extend their studies and obtain more generalizable results.

We will continue to promote data openness and sharing and actively encourage researchers from various fields to join the important work of cryptocurrency anti-money laundering, which benefits the healthy development of the blockchain ecosystem.

\section{Conclusion}
The lack of effective real-world address datasets on blockchains has hindered the progress of cybercrime detection and limited participation from researchers without cryptocurrency expertise. In this paper, we introduce the Real-CATS dataset, a comprehensive collection of labeled addresses from real-world reports across two blockchains. The dataset demonstrates the C3R characteristics, supporting three key functions to advance cybercrime detection research. We believe that the C3R framework and collection methodology can be extended to other blockchains, such as Polygon and Solana. By improving these datasets, we aim to help data science researchers from diverse fields contribute to anti-cybercrime efforts, develop better detection methods, enhance anti-money laundering research, and assist exchanges in meeting regulatory requirements, supporting the blockchain ecosystem's growth.

\section*{Ethics Considerations}
We carefully considered the ethical implications of our work while collecting datasets. It is important to emphasize that our research does not address privacy concerns. First, the data we collect is sourced from publicly available blockchain ledgers or databases, which can be accessed by anyone. Second, the data collection process utilizes anonymous blockchain transaction data, with no exposure to any actual personal information. Furthermore, our research aims to promote the healthy development of the cryptocurrency economy, uphold fairness and justice in transactions, and contribute to the regulation of cryptocurrency abuse by various countries, thereby facilitating the formulation of legal frameworks. This also has a positive impact on the protection of the public’s cryptocurrency assets.

\section*{Open Science}
All datasets proposed by this paper are open-sourced at \url{https://github.com/sjdseu/Real-CATS}. It includes 7 files related to the Real-CATS dataset and 1 file related to the Sup-CATS dataset. Additionally, we provide detailed transaction records for each address.

{\footnotesize \bibliographystyle{acm}
\bibliography{sample}}

\end{document}